\documentclass[aps,superscriptaddress,showpacs,nofootinbib,floatfix,epfs]{revtex4}
\pdfoutput=1
\usepackage{amsfonts,amscd,amsmath, amssymb,graphicx,color}

\def\W{W_{\text{\tiny gen}}}
\def\f0{f_{\text{\tiny 0}}}
\newcommand{\oh}{\frac{1}{2}}
\def\z{z_{\text{\tiny 0}}}
\def\ep{\text{e}}
\def\rt{r_{\text{\tiny T}}}

\def\g{\mathfrak{g}}
\def\oh{\frac{1}{2}}
\def\x0{x_{\text{\tiny 0}}}
\def\y0{y_{\text{\tiny 0}}}
\def\r0{r_{\text{\tiny 0}}}
\def\z0{z_{\text{\tiny 0}}}
\def\V{\mathbb{V}}
\def\hV{\tilde{\mathbb{V}}}
\def\hA{\mathbb{A}}
\def\m{\mathfrak{m}}
\def\s{\mathfrak{s}}
\def\last{\lambda_\ast}
\def\h{\text{e}^{\s r^2}}

\begin{document}
\title{On Exotic Hybrid Pseudo-Potentials and Gauge/String Duality}
\author{Oleg Andreev}
 \affiliation{L.D. Landau Institute for Theoretical Physics, Kosygina 2, 119334 Moscow, Russia}
\affiliation{Arnold Sommerfeld Center for Theoretical Physics, LMU-M\"unchen, Theresienstrasse 37, 80333 M\"unchen, Germany}
\begin{abstract} 
We use gauge/string duality to model some hybrid pseudo-potentials in a pure $SU(N)$ gauge theory. The pseudo-potentials under 
consideration can't be described by a single Nambu-Goto string. This is why we call them "exotic". A comparison 
with the calculation of the pseudo-potential shows the universality of the spatial string tension. 

\end{abstract}
\preprint{LMU-ASC 78/12}
\maketitle

\vspace{-7.5cm}
\begin{flushright}
LMU-ASC 78/12
\end{flushright}
\vspace{6cm}

\section{Introduction}
\renewcommand{\theequation}{1.\arabic{equation}}
\setcounter{equation}{0}

At high temperature, a pure $SU(N)$ gauge theory undergoes a phase transition from a confining phase to a phase which is deconfining in nature. While a perturbative analysis is appropriate at very high temperatures, where the physics is that of an asymptotically free gas of gluons, this is not the case at lower temperatures, where non-perturbative effects show up and make the perturbative analysis fail. 

One of the impacts of the non-perturbative physics arising from the magnetic sector is that the pseudo-potential extracted from spatially oriented 
Wilson loops is linear at large distances \cite{sW}. An important example to which this applies is a rectangular Wilson loop of size $R\times Y$, oriented in the $xy$-plane. In this case the expectation value of the loop is

\begin{equation}\label{sloop}
\langle W(R,Y)\rangle=\sum_{n=0}^{\infty} c_n(R)\,\ep^{-\V_n(R)Y}
\,.
\end{equation}
$\V$ is called the pseudo-potential if the corresponding contribution dominates the sum as $Y\rightarrow\infty$. If so, then the other $\V_n$'s are called hybrid pseudo-potentials. 

The pseudo-potential at finite temperature has been intensively studied in the past decades. In particular, the temperature dependence of the spatial string tension $\sigma_s$ was calculated on the lattice for $SU(2)$ and $SU(3)$ gauge groups \cite{st-lattice}. This showed the expected dependence on the magnetic scale and also established a relation to the string tension of a $3$-dimensional gauge theory. The latter is essential for another approach based on a dimensionally reduced gauge theory.\footnote{For details and references, see \cite{laine}.} 

On the other hand, nothing is known on the hybrid pseudo-potentials $\V_n$ at finite temperature. The situation here is quite different from that of the hybrid potentials, in that the hybrid potentials were calculated from the generalized temporal Wilson loops on the lattice as well as from four-dimensional effective string models \cite{kuti-rev}. An alternative route to studying these objects is an effective string theory in higher dimensions which is motivated by the AdS/CFT correspondence. 

In this paper we continue a series of studies \cite{az2,abar} devoted to the spatial Wilson loops within a five(ten)-dimensional effective string theory.\footnote{See also \cite{list}.} In \cite{az2}, the temperature dependence of the spatial string tension of an $SU(N)$ gauge theory without quarks was computed. For $T\lesssim 3T_c$ the result is remarkably consistent with the lattice data for $N = 2,\,3$ \cite{st-lattice}. Subsequent work showed that the multiquark pseudo-potentials obey the $Y$-law, and that the spatial string tension is universal \cite{abar}. We would like to emphasize that in these studies we don't need additional parameters than those used for computing the quark-antiquark and multiquark potentials in \cite{abar,az1}. 

Recently, we have proposed a construction of some hybrid potentials that perfectly reproduces the hybrid potential $\Sigma_u^-$ of \cite{kuti}. Here the key role is played by a new object called "defect" that is nothing but a macroscopic description of some gluonic degrees of freedom in strong coupling \cite{ahyb}. This provides a strong motivation for extending this construction to the hybrid pseudo-potentials. In doing so, we present here an example of the hybrid pseudo-potential $\Sigma_u^-$. At zero temperature it reduces to the corresponding hybrid potential of \cite{kuti}, as it should be. To our knowledge, this is the first example of a computation of a hybrid pseudo-potential in the literature.

The paper is organized as follows. In section II, we discuss the five(ten)-dimensional effective string theory. We begin by summarizing the theoretical background and the results of \cite{az2} for the pseudo-potential. Then, we introduce a point-like defect to macroscopically describe the $\Sigma$ hybrid pseudo-potentials. We go on in Sec.III to discuss the implications for a pure $SU(3)$ gauge theory. We conclude in Sec.IV with a brief discussion of possibilities for further study.

\section{Some Hybrid Pseudo-Potentials via Gauge/string Duality} 
\renewcommand{\theequation}{2.\arabic{equation}}
\setcounter{equation}{0}
\subsection{General formalism}

First, let us set the basic framework. The background metric in question is a one-parameter deformation 
of the Schwarzschild black hole in $\text{AdS}_5$ space \cite{az2}

\begin{equation}\label{metric}
ds^2={\cal R}^2 w \Bigl(fdt^2+d\vec x^2+f^{-1}dr^2\Bigr)\,,
\quad w(r)=\frac{\h}{r^2}
\,,
\quad f(r)=1-\Bigl(\frac{r}{\rt}\Bigr)^4
\,,
\end{equation}
where $d\vec x^2=dx^2+dy^2+dz^2$, $\cal R$ is a radius of AdS space, and $\s$ is a deformation parameter. $\rt$ is related to the Hawking temperature 
of the black hole, whose dual description is nothing but the temperature of gauge theory, as follows $\rt=1/\pi T$.  In addition, we take a constant dilaton and discard other background fields.

As a prelude to discussing the hybrid pseudo-potentials, let us briefly consider the pseudo-potential. As noted above, it is determined from the expectation value of a spatial Wilson loop. On the string theory side, we calculate expectation values of spatial Wilson loops by adopting the proposal of \cite{malda1}

\begin{equation}\label{wilson}
	\langle W({\cal C})\rangle\sim \ep^{-S}
	\,,
\end{equation}
where $S$ is an area of a string worldsheet bounded by a curve ${\cal C}$ at the boundary of AdS space. For the 
background geometry \eqref{metric}, the pseudo-potential $\V(R)$ extracted from a rectangular loop of size $R\times Y$ is written in parametric form as \cite{az2}

\begin{equation}\label{r}
R(\lambda)=2\sqrt{\frac{\lambda}{\s}}
\int_0^1 dv\,v^2\ep^{\lambda(1-v^2)}
\biggl(1-\Bigl(\frac{\lambda}{\tau}\Bigr)^2v^4\biggr)^{-\oh}
\biggl(1-v^4\ep^{2\lambda(1-v^2)}\biggr)^{-\oh}
\,,
\end{equation}

\begin{equation}\label{v0}
\V (\lambda)=2\g\sqrt{\frac{\s}{\lambda}}
\int_0^1 \frac{dv}{v^2}\Biggl[\ep^{\lambda v^2}
\biggl(1-\Bigl(\frac{\lambda}{\tau}\Bigr)^2v^4\biggr)^{-\oh}
\biggl(1-v^4\ep^{2\lambda (1-v^2)}\biggr)^{-\oh}-1-v^2\Biggr] + C
\,,
\end{equation}
where $\lambda$ is a parameter, $C$ is a normalization constant, $\tau=\s\rt^2$, and $\g=\frac{{\cal R}^2}{2\pi\alpha'}$. The parameter $\lambda$ takes values in the interval $[0,1]$ if $\tau\geq 1$ and $[0,\tau]$ if $\tau<1$.

At long distances the pseudo-potential is linear at any finite temperature

\begin{equation}\label{lr}
	\V(R)=\sigma_sR +C+o(1)
	\,,
\end{equation}
with 
\begin{equation}\label{st}
	\sigma_s=
	\begin{cases}
		\sigma\quad&\text{if}\quad T\leq T_c\,,\\
		\sigma \Bigl(\frac{T}{T_c}\Bigr)^2\exp\Bigl\{\Bigl(\frac{T_c}{T}\Bigr)^2-1\Bigr\}\quad&\text{if}\quad T> T_c
\end{cases}
\end{equation}
a spatial string tension. At this stage, we set $\sigma=e\g\s$ and $T_c=\sqrt{\s}/\pi$. Note that $\sigma$ is the physical string tension at zero temperature and $\tau=(T_c/T)^2$.

Obviously, there exists a critical value of $T$, such that the spatial string tension is temperature independent below $T_c$ and rises rapidly above. On the lattice such a pattern was found in \cite{st-lattice}. For $T\lesssim 3T_c$, the result \eqref{st} is remarkably consistent with the available 
lattice data for $SU(2)$ and $SU(3)$ gauge theories \cite{az2}. Note that at higher temperatures the temperature dependence of $\sigma_s$ 
is determined by the $\beta$-function of a gauge theory \cite{st-lattice}. Certainly, the model does not incorporate the running coupling that makes it 
fail in this temperature range.

At short distances the pseudo-potential is a power function of $R$ for any temperature

\begin{equation}\label{sr}
	\V(R)=-\frac{\alpha}{R}+C+o(1)
	\,,
\end{equation}
where $\alpha=(2\pi)^3\g/\Gamma^4(\frac{1}{4})$. This is another indication that the model is not perfect. On dimensional grounds, one expects that 
$\V$ is a logarithmic function of $R$ at very high temperatures.

\subsection{Exotic hybrids}
It is well known that the quark-antiquark potential corresponds to the ground state of a single string stretched between two fermionic sources. The common 
wisdom is that excited strings (fluxes) lead to hybrid potentials. On the lattice the corresponding operators are defined as linear combinations of the path-ordered exponentials of the gauge field \cite{kuti-rev}. The generalized Wilson loops are then constructed by parallel transporting these 
operators along the $t$-axis. However, making a transformation along a spatial direction introduces new objects. It is natural to call such objects 
generalized spatial Wilson loops. Hybrid pseudo-potentials are then extracted from the generalized spatial Wilson loops in the same way as the hybrid potentials are extracted from the generalized Wilson loops. It is clear from the construction that the hybrid pseudo-potentials may be labeled by the same set of quantum numbers. It also follows that given a string configuration for a hybrid potential, we can use it to construct the corresponding hybrid pseudo-potential, and vice versa. 

Recently, a string configuration for the $\Sigma$ hybrid potentials has been proposed in \cite{ahyb}. According to this proposal, some string excitations can be modeled by inserting local objects on a string. These objects are called defects. The reason for this is that in the presence of defects string embeddings of worldsheets into spacetime are not differentiable at points where defects are located. 

We will now attempt to use the configuration of \cite{ahyb} for finding the corresponding hybrid pseudo-potentials. In the case of a single defect, we 
consider the configuration shown in Fig.1. We place the quark-antiquark pair at the boundary points of the five(ten)-dimensional space such that each fermionic source is the endpoint of a fundamental string. The strings join at the defect in the interior.
\begin{figure}[htbp]
\centering
\includegraphics[width=5.75cm]{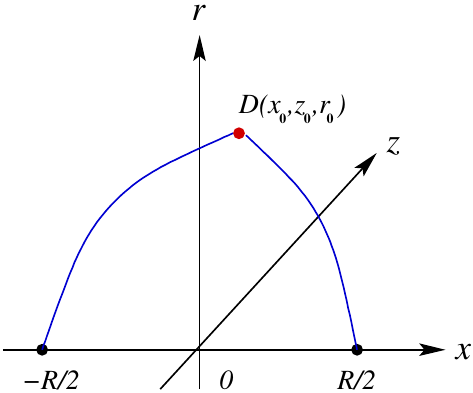}
\caption{{\small A configuration used to calculate the hybrid potentials. The quark and antiquark are set at $x=R/2$ and $x=-R/2$, respectively. The defect is placed at $D$.}}
\end{figure}
Thus, the action has in addition to the standard Nambu-Goto actions, also a contribution arising 
from the defect. It is given by  

\begin{equation}\label{stringaction}
S=\sum_{i=1}^2  S_i +S_{\text{\tiny def}}
\,,
\end{equation}
where $S_i$ denotes the action of the string connecting the $i$-source with the defect.

On the string theory side, a natural proposal for the expectation value of the generalized spatial Wilson loop is 

\begin{equation}\label{wloop-ads}
\langle\,\W ({\cal C})\,\rangle\sim \ep^{-S_{\text{\tiny min}}}
\,,
\end{equation}
where $S_{\text{\tiny min}}$ is an area of two string worldsheets bounded by a loop ${\cal C}$ at the boundary of AdS space and glued along a line in 
the bulk. In the large $Y$ limit $S_{\text{\tiny min}}=Y\hV$ such that a dominant contribution comes from a lateral surface
whose area is proportional to $Y$, while two end-surfaces only provide small corrections.

Since we are interested in a static configuration, for the strings we take 
 
\begin{equation}\label{gauge}
y_i(\tau_i)=\tau_i\,,\quad
x_i(\sigma_i)=a_i\sigma_i+b_i
\,.
\end{equation}
Here $(\tau_i,\sigma_i)$ are worldsheet coordinates. The Nambu-Goto action of the $i$-string is then

\begin{equation}\label{ng-action}
S_i=Y\g\int_0^1 d\sigma_i\,w\sqrt{a^2_i+ z'^2_i+f^{-1}r'^2_i}
\,,
\end{equation}
where a prime denotes a derivative with respect to $\sigma_i$.

The action for the defect is taken to be of the form

\begin{equation}\label{vertex}
S_{\text{\tiny def.}}=Y{\cal V}(\r0 )
\,,
\end{equation}
where ${\cal V}$ can be considered as its effective potential. Unfortunately, the explicit form of ${\cal V}$ is not determined only 
from the 5-dimensional metric. It requires the knowledge of the string theory dual to QCD. We will return to this issue in the 
next sections.

The boundary conditions on the fields are given by 

\begin{align}\label{boundary}
x_1(0)&=-R/2\,, 
&
x_1(1)&=\x0\,,
&
z_1(0)&=0\,,
&
z_1(1)&=\z0\,,
&
r_1(0)&=0\,,
&
r_1(1)&=\r0\,,\\
x_2(0)&=\x0\,, 
&
x_2(1)&=R/2\,,
&
z_2(0)&=\z0\,,
&
z_2(1)&=0\,,
&
r_2(0)&=\r0\,,
&
r_2(1)&=0\,.
\end{align}
These determine the coefficients $a_i$  and $b_i$ in \eqref{gauge}. Thus, we have

\begin{equation}\label{y}
a_1=\x0+R/2\,,\quad b_1=-R/2\,,
\quad
a_2=R/2-\x0\,,
\quad
b_2=\x0
\,.
\end{equation}

Now, we extremize the total action $S$ with respect to the worldsheet fields $z_i(\sigma_i)$ and 
$r_i(\sigma_i)$ describing the strings as well as with respect to $\x0$, $\z0$ and $\r0$ describing the location of the defect, with the 
following identifications: $\delta z_1(1)=\delta z_2(0)=\delta\z0$ and $\delta r_1(1)=\delta r_2(0)=\delta\r0$. In doing so, we use the fact that 
there are two symmetries which simplify the further analysis.

Since the integrand in \eqref{ng-action} does not depend explicitly on $\sigma_i$, we get the first integral of Euler-Lagrange equations

\begin{equation}\label{1integral}
I_i=\frac{w_i}{\sqrt{a_i^2+z_i'^2+f^{-1}r_i'^2}}
\,.
\end{equation}
In addition, because of translational invariance along the $z$-direction, there is another first integral. Combining 
it with \eqref{1integral} gives

\begin{equation}\label{2integral}
P_i=z'_i\,.
\end{equation}
Together with the boundary conditions these equations determine the $z_i$'s

\begin{equation}\label{x}
z_1 (\sigma_1 )=\z0\sigma_1\,,\quad
z_2(\sigma_2)=-\z0\sigma_2+\z0
\,.
\end{equation}

Next, we extremize the action with respect to the location of the defect. After using \eqref{y} and \eqref{x}, we get 

\begin{equation}\label{xyr-eqs}
\left(\x0+R/2\right)I_1+\left(\x0-R/2\right)I_2=0
\,,
\quad
\z0\left(I_1+I_2\right)=0
\,,
\quad
r_1'(1)I_1-r_2'(0)I_2+\g^{-1}f_0{\cal V}'(\r0)=0
\,,
\end{equation}
where $f_0=f(\r0)$. From this it follows that $\z0=0$. As a result, the static configuration lies entirely in the $xr$-plane.

Now we introduce $k_1=\left(\frac{r_1'(1)}{\x0+R/2}\right)^2$ and $k_2=\left(\frac{r_2'(0)}{R/2-\x0}\right)^2$. This allows us to rewrite 
the first equation of \eqref{xyr-eqs} as 

\begin{equation}\label{xr}
	\frac{1}{\sqrt{1+f_0^{-1}k_1}}-\frac{1}{\sqrt{1+f_0^{-1}k_2}}=0\,,
	\end{equation}
where $f_0=f(\r0)$. Obviously, it yields the unique solution $k_1=k_2=k$. Combining this with \eqref{xyr-eqs}, we have

\begin{equation}\label{r0}
	\frac{2}{\sqrt{f_0^{-1}+k^{-1}}}+\frac{1}{\g}\frac{f_0}{w}{\cal V'}(\r0)=0
	\,.
\end{equation}

If we define the first integrals $I_1$ at $\sigma_1=1$ and $I_2$ at $\sigma_2=0$ and integrate over $[0,1]$ of 
$d\sigma_i$, then by virtue of \eqref{xr} we get

\begin{equation}\label{l-eqs}
R\pm 2\x0=2
\sqrt{\frac{\lambda}{\s(1+f_0^{-1}k)}}
\int_0^1 dv\,v^2\ep^{\lambda(1-v^2)}
\biggl(1-\Bigl(\frac{\lambda}{\tau}\Bigr)^2v^4\biggr)^{-\oh}
\biggl(1-\frac{1}{1+f_0^{-1}k}v^4\ep^{2\lambda(1-v^2)}\biggr)^{-\oh}
\,.
\end{equation}
Obviously, it is consistent only if $\x0=0$. As a result, we end up with the most symmetric configuration.

Now, we will compute the energy of the configuration. First, we reduce the integrals over $\sigma_i$ in Eq.\eqref{ng-action} to that 
over $r_i$. This is easily done by using the first integral \eqref{1integral}. Since the integral is divergent at $r_i=0$, we regularize it by imposing 
a  cutoff $\epsilon$.\footnote{Importantly, in this process we use the same renormalization scheme as that for $\V$ in \cite{az2}.} 
Finally, the regularized expression takes the form

\begin{equation}\label{energy}
\hV_R=
{\cal V}(\lambda)+
2\g\sqrt{\frac{\s}{\lambda}}
\int_{\sqrt{\frac{\s}{\lambda}}\epsilon}^1 \frac{dv}{v^2}\, \ep^{\lambda v^2}
\biggl(1-\Bigl(\frac{\lambda}{\tau}\Bigr)^2v^4\biggr)^{-\oh}
\biggl(1-\frac{1}{1+f_0^{-1}k}v^4\ep^{2\lambda(1-v^2)}\biggr)^{-\oh}
\,.
\end{equation}
Its $\epsilon$-expansion is 

\begin{equation*}\label{energy2}
\hV_R=\frac{2\g}{\epsilon}+O(1)
\,.
\end{equation*}
Subtracting the $\frac{1}{\epsilon}$ term (quark masses) and letting $\epsilon=0$, we get a finite result

\begin{equation}\label{energy3}
\hV(\lambda)={\cal V}(\lambda)
+2\g\sqrt{\frac{\s}{\lambda}}
\int_0^1 \frac{dv}{v^2}\Biggl[\ep^{\lambda v^2}
\biggl(1-\Bigl(\frac{\lambda}{\tau}\Bigr)^2v^4\biggr)^{-\oh}
\biggl(1-\frac{1}{1+f_0^{-1}k}v^4\ep^{2\lambda (1-v^2)}\biggr)^{-\oh}-1-v^2\Biggr] + C
\,,
\end{equation}
where $C$ is the same normalization constant as in \eqref{v0}. Note that \eqref{energy3} reduces to \eqref{v0} at ${\cal V}=0$, as promised in the 
introduction.\footnote{It follows from \eqref{r0} that $k=0$ at ${\cal V}=0$.}

\subsection{Concrete examples}

In this section, we will describe a couple of concrete examples in which one can develop a level of understanding which is similar to that of the exotic hybrid potentials \cite{ahyb}. 

\subsubsection{Model A}

Following \cite{ahyb}, we first consider the five-dimensional geometry \eqref{metric}. In this case the action $S_{\text{\tiny def}}$ is that of a 
point like particle $\m\int ds$, with $\m$ a particle mass. The latter implies that 

\begin{equation}\label{potential}
{\cal V}(\r0 )=\m{\cal R}\,\ep^{\oh\s\r0^2}/\r0
\,.
\end{equation}

With the help of \eqref{potential}, we can compute $R$ and $\hV$ as functions of $\lambda$, getting

\begin{equation}\label{l-eqs-ex}
R(\lambda)=2\sqrt{\frac{\lambda}{\s}\rho}\int_0^1 dv\,v^2\ep^{\lambda (1-v^2)}
\Bigl(1-\Bigl(\frac{\lambda}{\tau}\Bigr)^2v^4\Bigr)^{-\oh}
\Bigl(1-\rho \,v^4\ep^{2\lambda (1-v^2)}\Bigr)^{-\oh}
\,
\end{equation}
and
\begin{equation}\label{energy-ex}
\hV(\lambda)=2\g\sqrt{\frac{\s}{\lambda}}
\biggl(\kappa\,\ep^{\oh\lambda}-1
+
\int_0^1 \frac{dv}{v^2}\biggl[\ep^{\lambda v^2}
\Bigl(1-\Bigl(\frac{\lambda}{\tau}\Bigr)^2v^4\Bigr)^{-\oh}
\Bigl(1-\rho \,v^4\ep^{2\lambda (1-v^2)}\Bigr)^{-\oh}-1\biggr]\biggr)+C
\,,
\end{equation}
where $\kappa=\frac{\m{\cal R}}{2\g}$ and $\rho(\lambda)=1-\kappa^2(1-(\lambda/\tau)^2)(1-\lambda)^2 \ep^{-\lambda}$. For applications to realistic gauge theories, $\kappa$ must be larger than $1$ \cite{ahyb}. In this case, a simple analysis shows that the parameter $\lambda$ takes values in the interval $[\last,\min(1,\tau)]$, where $\last$ is a solution of equation $\rho(\lambda)=0$.

As in the case of the hybrid potential \cite{ahyb}, the hybrid pseudo-potential is written in parametric form. It is unclear to us how to eliminate the parameter $\lambda$ and find $\hV$ as a function of $R$. We can, however, gain some important insights from numerical calculations. In Fig.2 on the left, we plot $\hV$ against $R$. We see that at large distances the hybrid pseudo-potential is always linear. Moreover, its slope depends only on $T$ but not on $\kappa$. 
\begin{figure}[htbp]
\centering
\includegraphics[width=5.5cm]{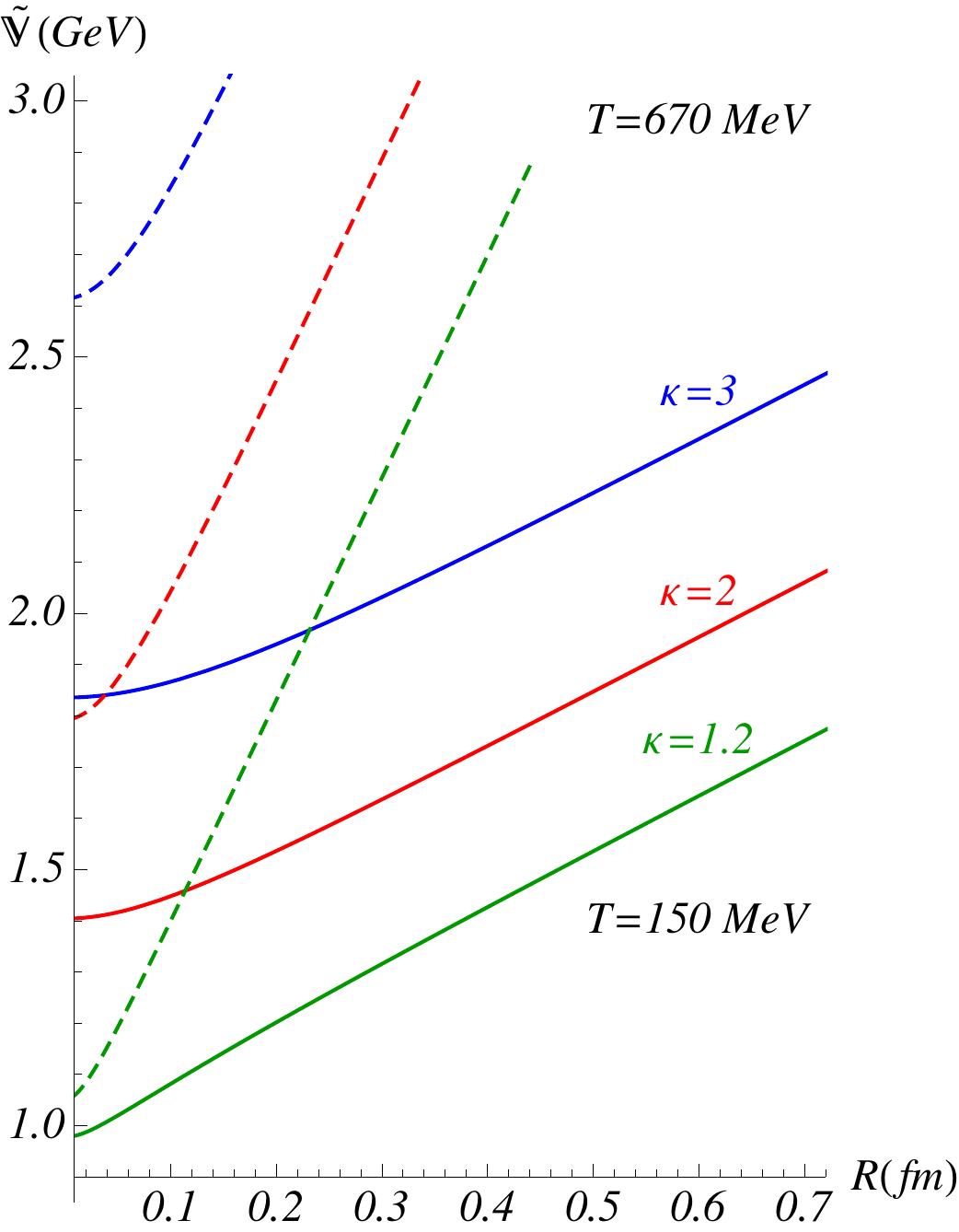}
\hspace{4cm}
\includegraphics[width=5.5cm]{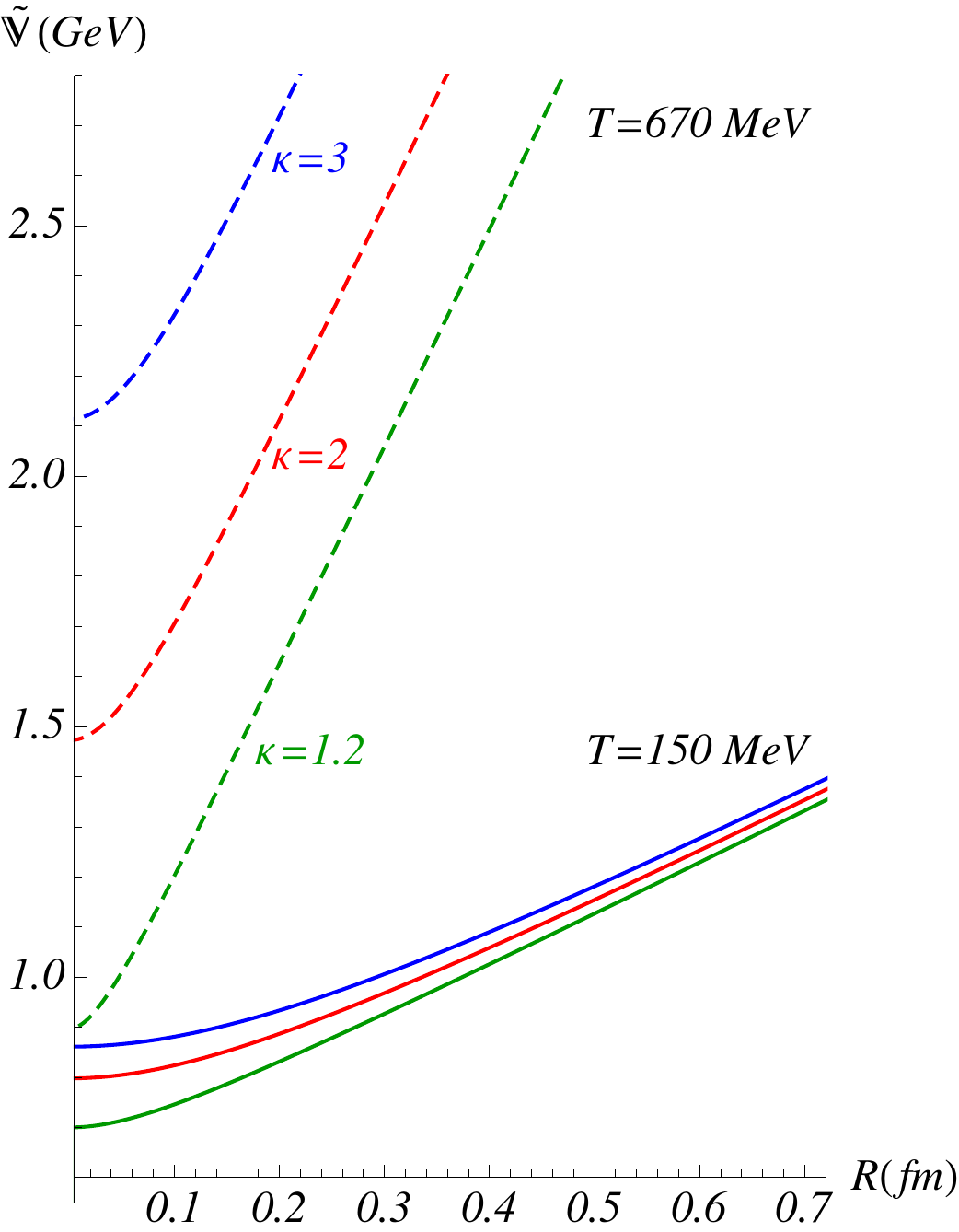}
\caption{{\small $\hV$ as a function of $R$ at $T=150,\, 670\,\text{MeV}$ and $\kappa=1.2,\, 2,\, 3$. Left: Model A. Right: Model B.
In both cases, $\g=0.176$, $\s=0.44\,\text{GeV}^2$, and $C=0.71\,\text{GeV}$, as in \cite{ahyb}.}}
\end{figure}
The picture changes at short distances, where the form of the hybrid pseudo-potential becomes dependent on the value of $\kappa$ as well.

Having seen the pattern from the numerics, we can try to solve the problem in the two limiting cases, long and short distances.

At long distances, the hybrid pseudo-potential $\hV(R)$ is linear for any finite value of $\kappa$

\begin{equation}\label{largerA}
	\hV(R)=\sigma_sR+C+\Delta_s+o(1)
\,,
\end{equation} 
with $\sigma_s$ being the spatial string tension \eqref{st}. This suggests that the spatial string tension is universal. From the 5d perspective, the system is prevented from getting deeper into the $r$-direction than the soft wall ($r=1/\sqrt{\s}$) at low temperatures and the horizon ($r=\rt$) at high temperatures. The phase transition point is at $\rt=1/\sqrt{\s}$ or, equivalently, at $\tau=1$ \cite{az2}.

The point here is that there exists a finite gap between $\hV$ and $\V$ 

\begin{equation}\label{gap}
\Delta_s=\lim_{R\to\infty}\hV(R)-\V(R)=2\kappa\sqrt{\g\sigma_s}
\,,
\end{equation}
which follows from the expression \eqref{potential} evaluated at $\lambda=\{1,\tau\}$. It is noteworthy that the gap doesn't show any 
temperature dependence up to $T_c$ but then it rises like $\sqrt{\sigma_s}$.

At short distances, the hybrid pseudo-potential is given by 

\begin{equation}\label{smallrA3}
	\hV(R)= \hV(\last)+\hA(\last) R^2+O(R^3)
	\,,
\end{equation}
where $\hV(\last)$ is a value of \eqref{energy-ex} at $\lambda=\last$ 
\begin{equation}\label{tV}
\hV(\last)=2\g\sqrt{\frac{\s}{\last}}\biggl(\kappa\,\ep^{\oh\last}-1+
\int_0^1 \frac{dv}{v^2}\biggl[\ep^{\last v^2}
\Bigl(1-\Bigl(\frac{\last}{\tau}\Bigr)^2v^4\Bigr)^{-\oh}-1\biggr]\biggr)
+C
\,
\end{equation}
and 
\begin{equation}\label{coefA}
\hA(\last)=\frac{\g}{4}\Bigl(\frac{\s}{\last}\Bigr)^{\frac{3}{2}}\ep^{\last}
\biggl(
\int_0^1 dv\, v^2 \ep^{\last (1- v^2)}\Bigl(1-\Bigl(\frac{\last}{\tau}\Bigr)^2v^4\Bigr)^{-\oh}
\biggr)^{-1}
\,.
\end{equation}
In these expressions, $\hV(\last)$ and $\hA(\last)$ are the functions of $T$ and $\kappa$ and, therefore, the asymptotic behavior of the pseudo-potential depends on both $T$ and $\kappa$, as also seen from Fig.2. Notice that $\hA$ is positive for all $\last$ of interest.

\subsubsection{Model B}

As in our study of the hybrid potentials, one can consider a defect living in ten-dimensional space \cite{ahyb}. For example, a flux loop is interpreted 
as a pair of baryon-antibaryon vertices connected by fundamental strings. According to \cite{edw}, in ten-dimensional space it is nothing but a brane-antibrane pair. Since the branes in question are fivebranes, it seems natural to assume that $S_{\text{\tiny def}}\sim\tau_5\int d^6x\sqrt{g^{(6)}}$, with $\tau_5$ a brane tension.

For subsequent applications, we will need to know the 10-dimensional geometry. One possibility here is to take \cite{a-pis} 

\begin{equation}\label{metric10}
ds^2={\cal R}^2 w \bigl(f dt^2+d\vec x^2+f^{-1}dr^2\bigr)+\text{e}^{-\s r^2}g_{ab}^{(5)}d\omega^a d\omega^b 
\,.
\end{equation}
This is a deformed product of $AdS_5$ and a 5-dimensional compact space (sphere) $\mathbf{X}$ whose coordinates are $\omega^a$. The deformation is due to the $r$-dependent warp factor. 

Because the fivebranes wrap on $\mathbf{R}\times\mathbf{X}$, with $\mathbf R$ along the $y$-axis in the deformed $\text{AdS}_5$, we have

\begin{equation}\label{potential10}
{\cal V}(\r0 )=\m{\cal R}\,\text{e}^{-2\s\r0^2}/\r0
\,.
\end{equation}
Note that $\m\sim\tau_5\int d^5\omega\sqrt{g^{(5)}}$.

We are now ready to reproduce the pseudo-potential. A short calculation gives

\begin{equation}\label{l-eqs-ex10}
	R(\lambda)=2\sqrt{\frac{\lambda}{\s}\bar\rho}\int_0^1 dv\,v^2\ep^{\lambda (1-v^2)}
	\Bigl(1-\Bigl(\frac{\lambda}{\tau}\Bigr)^2v^4\Bigr)^{-\oh}
	\Bigl(1-\bar\rho \,v^4\ep^{2\lambda (1-v^2)}\Bigr)^{-\oh}
	\,
\end{equation}
and
\begin{equation}\label{energy-ex10}
\hV(\lambda)=2\g\sqrt{\frac{\s}{\lambda}}
\biggl(\kappa\,\ep^{-2\lambda}-1
+
\int_0^1 \frac{dv}{v^2}\biggl[\ep^{\lambda v^2}\Bigl(1-\Bigl(\frac{\lambda}{\tau}\Bigr)^2v^4\Bigr)^{-\oh}
\Bigl(1-\bar\rho \,v^4\ep^{2\lambda (1-v^2)}\Bigr)^{-\oh}-1\biggr]\biggr)+C
\,,
\end{equation}
where $\bar\rho(\lambda)=1-\kappa^2(1-(\lambda/\tau)^2)\left(1+4\lambda\right)^2 \ep^{-6\lambda}$, with the same $\kappa$ as in \eqref{energy-ex}. The parameter $\lambda$ takes values in the 
interval $[\last,\lambda_c]$ if $\tau>1$ and $[\last,\tau]$ if $\tau\leq 1$. Here $\last$ is a solution of equation $\bar\rho(\lambda)=0$ and 
$\lambda_c$ is a solution of equation $\bar\rho(\lambda)=\lambda^2\ep^{2(1-\lambda)}$. Notice that $\lambda_c=1$ at $\tau=1$. 

Given the parametric form that we have just described, it is straightforward to gain some insights on the hybrid pseudo-potential from numerics. In Fig.2 on the right, we plot $\hV$ against $R$ for Model B. We see that the pattern is similar to that of Model A. 

As in our study of Model A, we can find explicit formulas in the two limiting cases, long and short distances. 

A simple calculation shows that the long distance behavior of $\hV$ is given by 

\begin{equation}\label{largerB}
	\hV(R)=\sigma_sR +C+\Delta_s+o(1)
	\,,
\end{equation}
with the same spatial string tension as in \eqref{lr}. Thus, Model B also suggests that the spatial string tension is universal. However, the expression 
for the finite gap $\Delta_s$ turns out to be more involved than \eqref{gap}. It is now 

\begin{equation}\label{gapB}
	\Delta_s= 2\sqrt{\ep\g\sigma}
	\begin{cases}
\kappa\lambda_c^{-\oh}\ep^{-2\lambda_c-1}+
	\int_1^{\sqrt{\lambda_c}}dv\bigl(v^{-4}\ep^{2(v^2-1)}-1\bigr)^{\oh}\bigl(1-(\tfrac{T}{T_c})^4v^4\bigr)^{-\oh}
	\quad&\text{if}\quad T < T_c\,,\\
\kappa\tfrac{T}{T_c}\exp\bigl\{-2\bigl(\tfrac{T_c}{T}\bigr)^2-1\bigr\}	\quad&\text{if}\quad T\geq T_c\,.
\end{cases}
\end{equation}
The meaning of this formula is as follows. At low temperatures, $T<T_c$, the first term is a contribution of the local defect. It is nothing but a value of the effective potential \eqref{potential10} at $\lambda=\lambda_c$. The second term represents a contribution of the string piece which gets beyond the soft wall ($r=1/\sqrt{\s}$) in the bulk. Here the point is that in Model B the system is allowed to go beyond the soft wall \cite{ahyb}. This makes it very different from single strings \cite{az1}, baryon vertices \cite{abar} and Model A, where it is not allowed to do so. Note that only a finite piece of the infinitely long string goes beyond the wall. This piece is not able to change the leading behavior at large $R$. But what it can do is modify the size of next-to-leading order corrections. At high temperatures, $T\geq T_c$, the large distance behavior is determined by the near horizon geometry. In this case the system is not allowed to go beyond the horizon. As a result, there is only one contribution to the gap which is due to the local defect. It is a value of \eqref{potential10} at $\lambda=\tau$.

At short distances, the pseudo-potential behaves as 

\begin{equation}\label{smallB}
	\hV(R)=\hV(\last)+\hA(\last)R^2+O(R^3)
	\,,
\end{equation}
where $\hV(\last)$ is a value of \eqref{energy-ex10} at $\lambda=\last$ 
\begin{equation}\label{VB}
\hV(\last)=2\g\sqrt{\frac{\s}{\last}}
\biggl(\kappa\,\ep^{-2\last}-1+
\int_0^1 \frac{dv}{v^2}\biggl[\ep^{\last v^2}
\Bigl(1-\Bigl(\frac{\last}{\tau}\Bigr)^2v^4\Bigr)^{-\oh}-1\biggr]\biggr)
+C
\,
\end{equation}
and $\hA$ as a function of $\last$ is given by \eqref{coefA}. 

\section{Implications for a pure SU(3) gauge theory}
\renewcommand{\theequation}{3.\arabic{equation}}
\setcounter{equation}{0}

So far we have dealt with the string theory construction in higher dimensions. However, the 
formalism developed can be used to predict properties of the $\Sigma$ hybrid pseudo-potentials in a pure $SU(3)$ gauge theory at finite temperature, in much the same way as it describes the the corresponding hybrid potentials at zero temperature \cite{ahyb}. 

In this section, we will be particularly interested in the $\Sigma_g^+$ and $\Sigma_u^-$ pseudo-potentials.\footnote{Note that in these notations $\Sigma_g^+$ stands for the string ground state, while $\Sigma_u^-$ is for the first $\Sigma$ excited state \cite{kuti-rev}.} For $\Sigma_g^+$ defined by \eqref{r} and \eqref{v0} there are three parameters: $\g$, $\s$, and $C$. Because these parameters are also used in the model of the heavy quark potential ($\Sigma_g^+$ potential) of \cite{az1}, we take the recent results of \cite{ahyb}, where the parameter values were fitted to the lattice data of \cite{kuti}. This yields $\g=0.176$, $\s=0.44\,\text{GeV}^2$, and $C=0.71\,\text{GeV}$. With these parameters fixed, the only parameter which remains to be determined for the $\Sigma_u^-$ hybrid pseudo-potential is $\kappa$. Again, we take the result of \cite{ahyb}, where it was fitted to the lattice data of \cite{kuti} for the $\Sigma_u^-$ potential. This yields $\kappa=2.3$ for Model A and $\kappa=2000$ for Model B. Thus, all the parameters are fixed from the $\Sigma$ potentials, and we don't need any free parameter to describe the $\Sigma$ pseudo-potentials. 

In Fig.3 on the left, we plot the pseudo-potentials at the physical  
\begin{figure}[htbp]
\centering
\includegraphics[width=5.75cm]{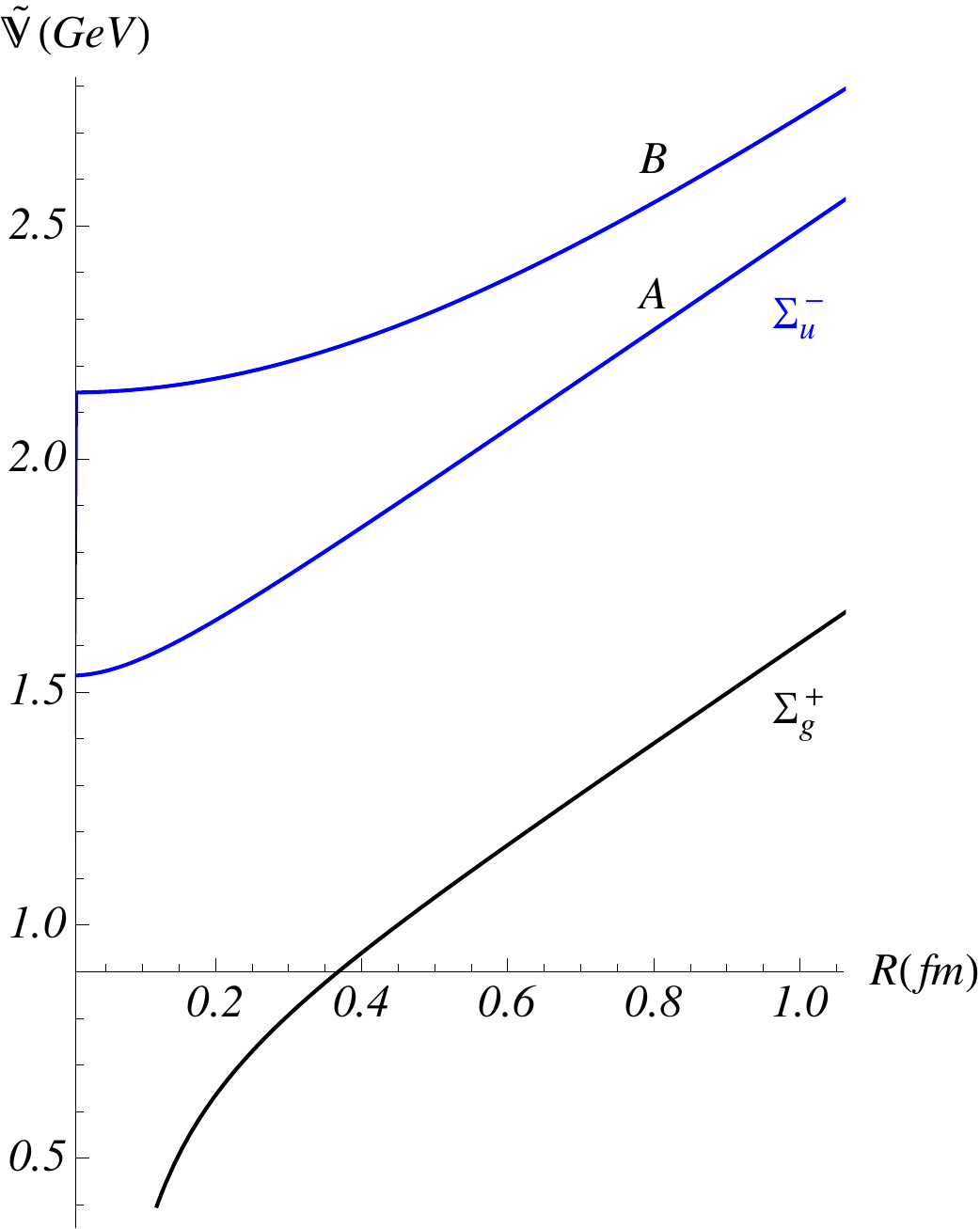}
\hspace{4cm}
\includegraphics[width=5.0cm]{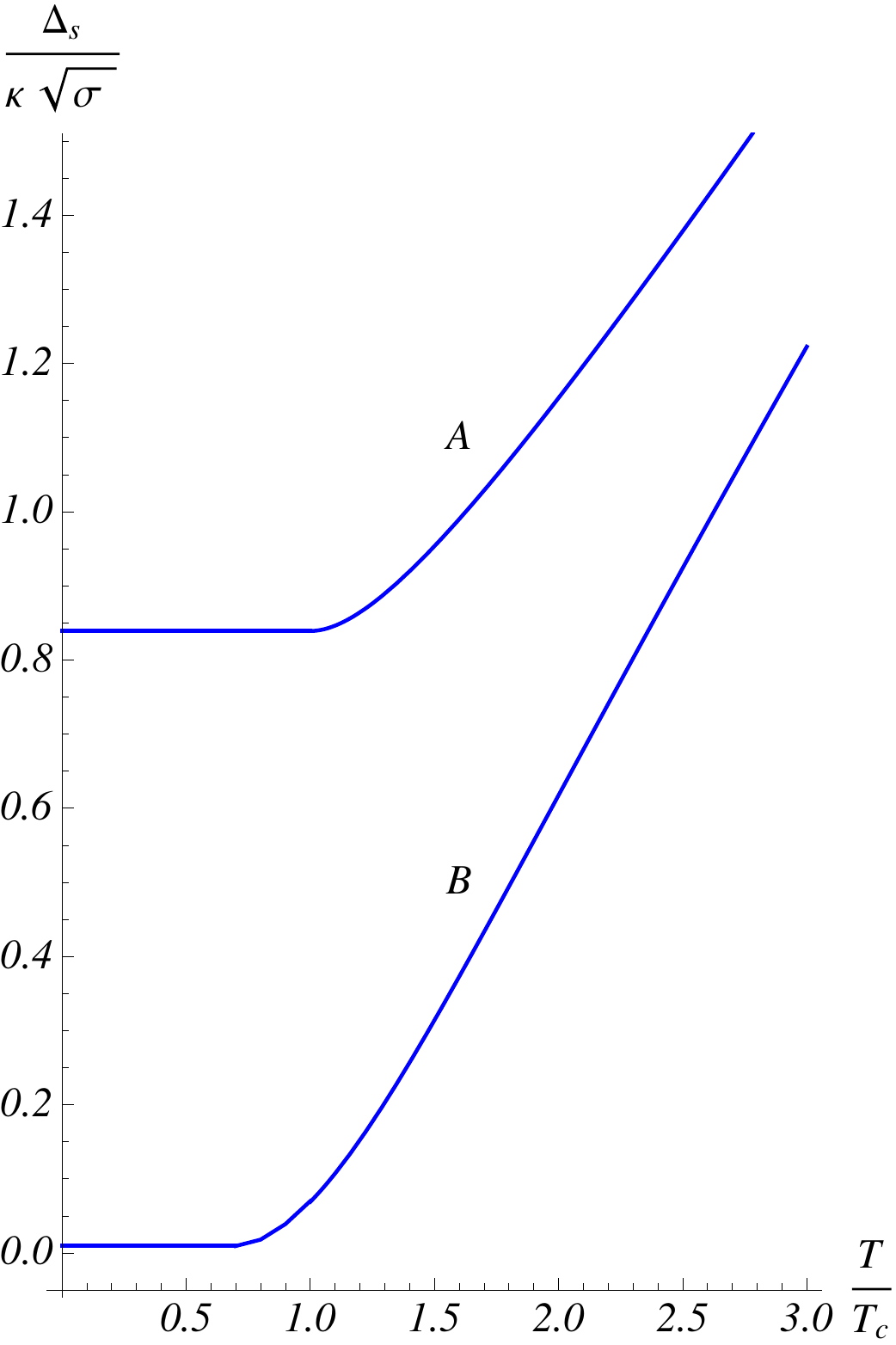}
\caption{{\small Left: The pseudo-potentials $\Sigma_g^+$ and $\Sigma_u^-$ at $T=0.5\,T_c$. Right: $\Delta_s/\kappa\sqrt{\sigma}$ as a function of $T/T_c$.}}
\end{figure}
parameter values. It follows immediately from the construction of $\hV$ that this reflects a general behavior pattern at finite temperature. Moreover, the long distance behavior of the pseudo-potentials is always linear with the same spatial string tension. We take this as an indication that all generalized spatial Wilson loops obey the area law behavior at any finite temperature and the spatial string tension is universal. It is noteworthy that in the case of the hybrid potentials Model B turned out to be superior to Model A as compared with the lattice data \cite{ahyb}. Therefore, we expect that this will be the case for the hybrid pseudo-potentials as well.

One of the interesting conclusions of our construction is that at large distances the $\Sigma_g^+$ and $\Sigma_u^-$ pseudo-potentials are separated by 
a finite gap $\Delta_s$. At zero temperature this gap coincides with the physical energy gap $\Delta$ between the $\Sigma_g^+$ and $\Sigma_u^-$ potentials \cite{ahyb}. In Fig.3 on the right, we plot the $\Delta_s/\kappa\sqrt{\sigma}$ obtained from Eqs.\eqref{gap} and \eqref{gapB} versus $T/T_c$. Here we restrict ourself to a temperature range of $0-3\,T_c$. As noted before, the reason is that the effective string theory description is no longer reliable 
for higher temperatures \cite{az2}, where asymptotic freedom allows perturbative predictions. Obviously, the gap doesn't show any significant temperature dependence at low temperatures. For Model A it is a constant function, while for Model B it is a slowly varying function of the temperature.  
The gap begins to grow when approaching the critical temperature.\footnote{Interestingly, in the case of Model A this happens exactly at $T=T_c$.} Note that in the temperature interval shown in Fig.3 the growth is not exactly linear.

Although our construction may be inappropriate at short distances, it is, however, worth seeing how everything looks and works.

In both models the hybrid pseudo-potential after extrapolation to small $R$ behaves as 

\begin{equation}\label{smallR}
	\hV(R)=\hV_0+\hA R^2+O(R^3)\,,
\end{equation}
with $\hV_0$ given by \eqref{tV}, \eqref{VB} and $\hA$ given by \eqref{coefA}. At zero temperature it coincides with that of the hybrid potential \cite{ahyb}. The constant term $\hV_0$ is scheme dependent. It includes the common renormalization constant $C$, such that in the $\Sigma_g^+$ pseudo-potential \eqref{v0}. On the other hand, the value of $\hA$ is scheme independent. 

Now let us see what happens to $\hV_0$ and $\hA$ at finite temperature. In Fig.4, we present the results for the temperature range of $0-3\,T_c$. 
We see that in both models the coefficients show a similar behavior. At low temperatures $\hV_0$ and $\hA$
\begin{figure}[htbp]
\centering
\includegraphics[width=5.25cm]{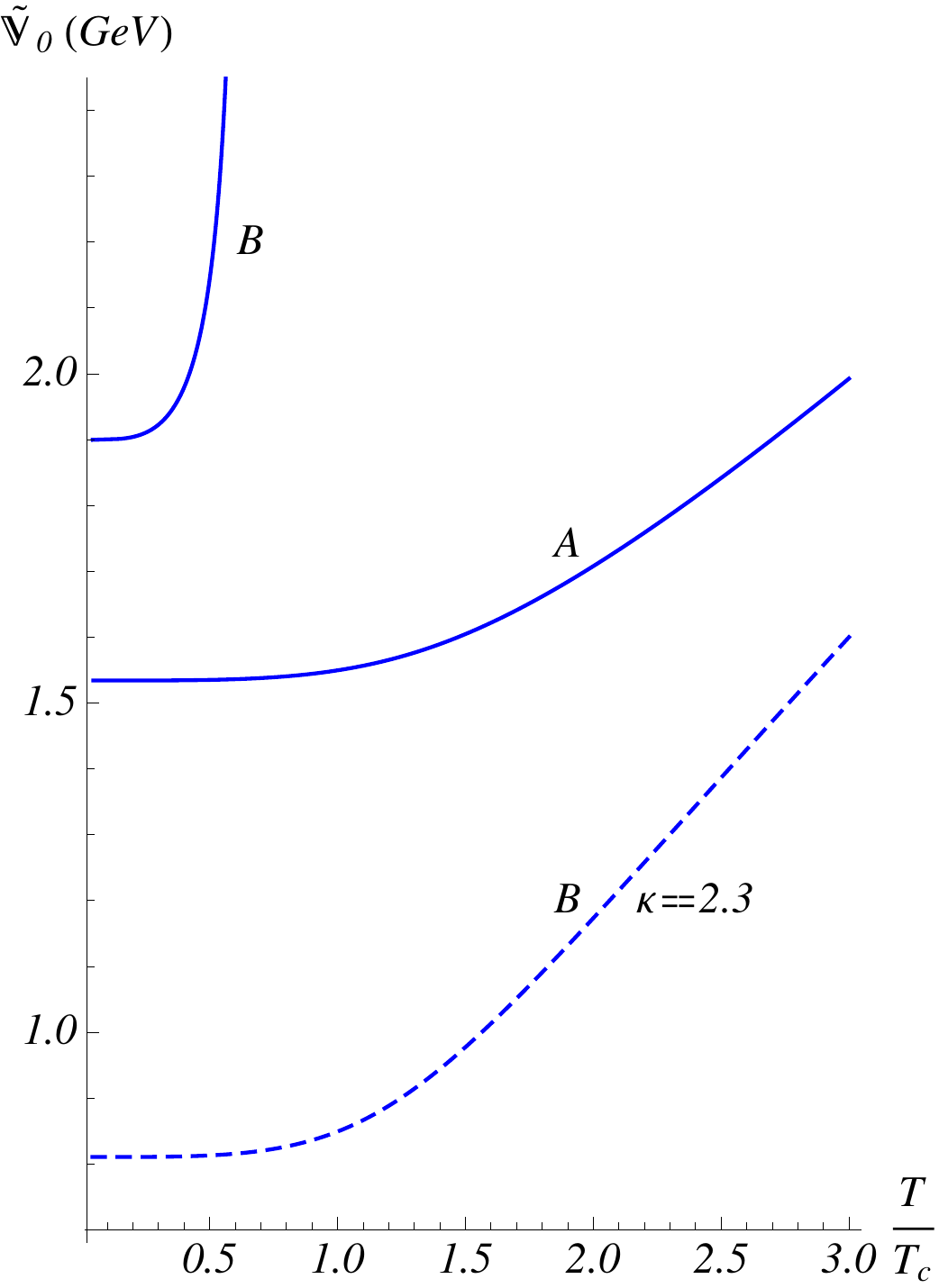}
\hspace{4cm}
\includegraphics[width=5.25cm]{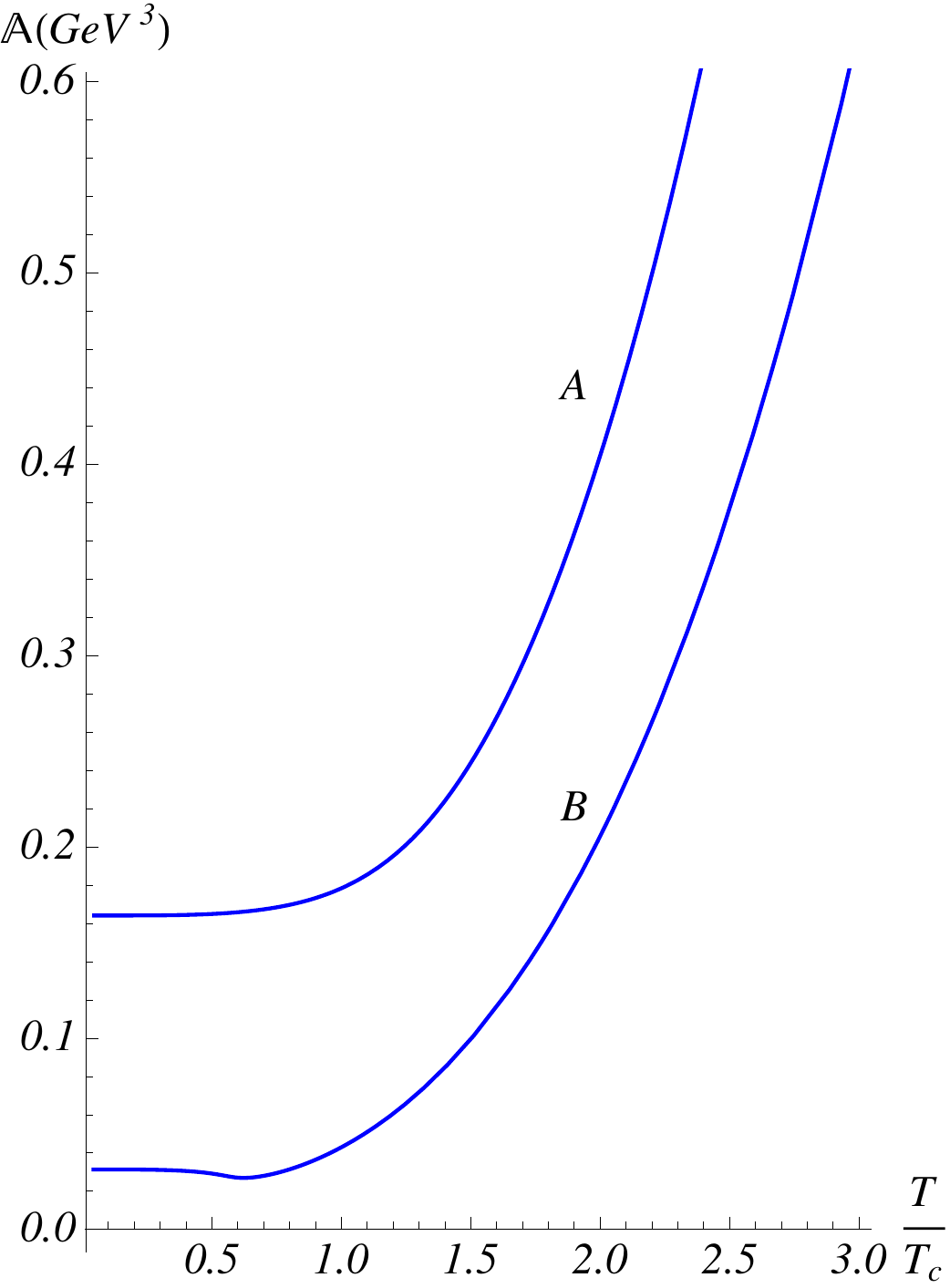}
\caption{{\small Coefficients $\hV_0$ and $\hA$ as functions of $T/T_c$.}}
\end{figure}
are slowly varying functions of $T$. As the temperature approaches the critical value, the functions begin to rise. 

We conclude our discussion of $\hV_0$ and $\hA$ by making a few remarks. First, $\hV_0$ is an increasing function of $\kappa$. For $\kappa$ of order $1$, 
Model A yields larger values of $\hV_0$ than Model B. This is illustrated in Fig.4 on the left and also seen from Fig.2. As the value of $\kappa$ in Model B is set to $2000$, the situation changes, as seen from Fig.4 on the left. In particular, this explains the relative positions of the curves in Fig.2 and Fig.3. Second, for $T\gtrsim 1.5\,T_c$ the curves for $\hA$ are fitted to the parametrization $\hA=a_0(T/T_c)^3+a_1$. Finally, in the case of Model B the function $\hA(T/T_c)$ has a shallow minimum at $T/T_c\approx 0.65$.

\section{Concluding Comments}
\renewcommand{\theequation}{4.\arabic{equation}}
\setcounter{equation}{0}

(i) In this work, we have restricted ourselves to the $\Sigma_g^+$ and $\Sigma_u^-$ pseudo-potentials. The reason for doing this is that for $R$ below about $0.6$ fm the lattice data for other hybrid potentials \cite{kuti} are well described by neither Model A nor Model B \cite{ahyb}. One might expect that the same thing also happens for the corresponding pseudo-potentials. It will be interesting, nevertheless, to see what actually happens. To our knowledge, there have been no studies of the hybrid pseudo-potentials on the lattice yet.  

The model we are pursuing predicts that the spatial string tension in units of the physical tension is universal in the sense that it depends only on the ratio $T/T_c$ (see Eq.\eqref{st}). This is supported by lattice simulations for $N=2,\,3$ \cite{az2}. However, there are no lattice data for larger $N$, and so new simulations would be welcome. The same is also applicable to the hybrid potentials whose $N$-dependence is not yet known.

(ii) Like in the case of zero temperature \cite{ahyb}, there are some important distinctions between static strings connecting the quark sources on the boundary of the deformed $\text{AdS}$ space, see Fig.5. 

\begin{figure}[htbp]
\centering
\includegraphics[width=6.25cm]{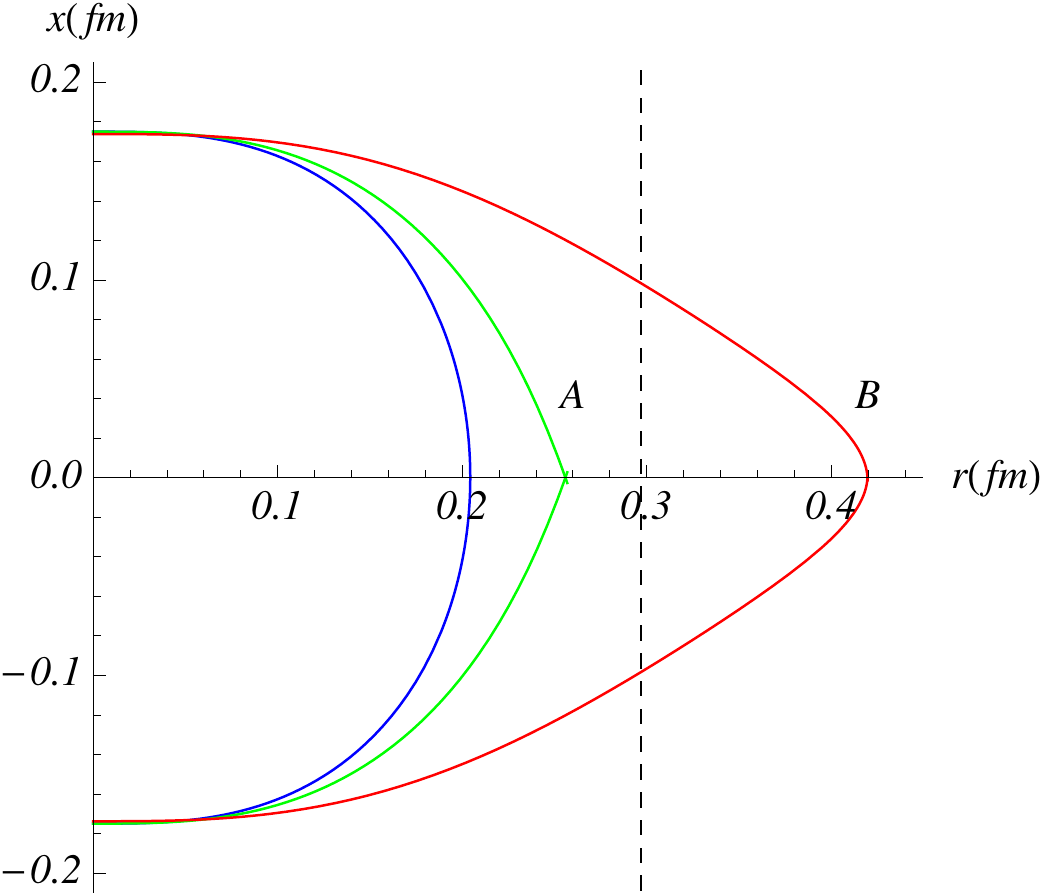}
\caption{{\small Static strings bend away from the boundary into the bulk. The dashed line indicates a soft wall. Here $\s=0.44\,\text{GeV}^2$ and $T=150\,\text{MeV}$.}}
\end{figure}

First, in Models A and B the defects lead to the formation of cusps at $x=0$. The cusp (deviation) angle is given by 

\begin{equation}\label{cusp}
\cos\frac{\theta(\lambda)}{2}=
\begin{cases}
	\Bigl(1+f_0(\rho^{-1}-1)\Bigr)^{-\oh}\\
	\Bigl(1+f_0(\bar\rho^{-1}-1)\Bigr)^{-\oh}
\,\,,
\end{cases}
\end{equation}
where $f_0=1-(\lambda/\tau)^2$. The upper expression holds in Model A and the lower expression holds in Model B. Note that in Fig.5 the cusp angle 
corresponding to Model B is very small. It is of order $0.01$.

Second, in Model A the cusp angle vanishes in the limit $R\rightarrow\infty$. This is true for all values of $T$. On the other hand, in Model B the cusp angle is non-zero for $T<T_c$, while it is zero for $T>T_c$. Thus, large strings are smoothed out in the deconfining phase. 

Third, in Model B the defect is allowed to go beyond the soft wall for low temperatures. However, its penetration depth decreases with temperature.

Finally, let us note that cusps occur in the five(ten)-dimensional framework, where the derivative $dr/dx$ turns out to be discontinuous. The question naturally arises: What happens in four dimensions? In other words, whether a QCD string is smooth or not.\footnote{We thank A.M. Polyakov for a discussion of this issue.}

\begin{acknowledgments}
We would like to thank Peter Weisz for helpful discussions and comments. 
We also thank the Arnold Sommerfeld Center for Theoretical Physics at LMU for its hospitality during much of this work. 
\end{acknowledgments}

\end{document}